%% file: main.tex
\documentclass[tighten, times,twocolumn]{aastex631}

\usepackage{epsfig}
\usepackage{amsmath,bm}
\usepackage{amssymb}
\usepackage{mathtools}
\usepackage{color}
\usepackage{natbib}
\usepackage{enumitem}
\usepackage{array}
\usepackage{makecell}
\usepackage{cleveref}
\usepackage{multirow}
\usepackage{cancel}
\usepackage[most]{tcolorbox}

\newcommand{\bvec}[1]{{\mbox{{\boldmath$#1$}}}}	
\newcommand{\unitv}[1]{\bvec{\hat{#1}}}			
\newcommand{\pdv}[2]{\frac{\partial{#1}}{\partial{#2}}}
\newcommand{\dv}[2]{\frac{d{#1}}{d{#2}}}

\input Figures.tex

\begin{document}

\title{Effects of density stratification on Rossby waves in deep atmospheres}

\author[0000-0001-9004-5963]{Catherine C. Blume}
\affiliation{JILA \& Department of Astrophysical and Planetary Sciences,
University of Colorado Boulder,
Boulder, CO 80309-0440, USA}

\author[0000-0001-7612-6628]{Bradley Hindman}
\affiliation{JILA \& Department of Astrophysical and Planetary Sciences,
University of Colorado Boulder,
Boulder, CO 80309-0440, USA}

\affiliation{Department of Applied Mathematics,
University of Colorado Boulder,
Boulder, CO 80309-0526, USA}

\begin{abstract}
{Though Rossby waves have been observed on the Sun, their radial eigenfunctions remain a mystery. Prior theoretical work either considers quasi-2D systems, which do not apply to the solar interior, or only considers fully radiative or fully convective atmospheres. Here we derive a radial wave equation that is valid for Rossby waves in a deep atmosphere with a general gravitational stratification. We adopt the $\beta$-plane approximation and express our equations in terms of the Lagrangian pressure fluctuation $\delta P$. We then calculate radial eigenfunctions for Rossby waves in a standard solar model, Model S. We find that working in the Lagrangian pressure fluctuation results in cleaner wave equations that lack the internal singularities that have been encountered in prior work. The resulting radial wave equation makes it abundantly clear that there are two wave cavities in the solar interior, one in the radiative interior and another in the convection zone. Surprisingly, our calculated radial vorticity eigenfunctions for the radiative interior modes are nearly constant throughout the convection zone, raising the possibility that they may be observable at the solar surface.}
\end{abstract}


\section{Introduction}

Rossby waves (also known as planetary waves or $r$ modes) are large-scale toroidal oscillations for which the restoring force is the Coriolis force. Though these waves have long been recognized as a dynamically important transport mechanism in Earth's atmosphere and oceans \citep[][]{Rossby1939RelationBV,1940TrAGU..21..262H}, they were only recently observed on the solar surface \citep[e.g.,][]{Loptien2018, 2019A&A...626A...3L,Proxauf2020,Alshehhi2019, Hanasoge2019,Hanson2020,2021AA...652L...6G,Hathaway2021, Waidele2023,Hanson2024}. An ongoing hope is that these modes will prove to be an excellent seismic diagnostic of properties of the solar interior that are currently poorly constrained (such as the convection zone's superadiabatic gradient). Before Rossby waves can be exploited in such a way, we need to understand how they propagate radially through different kinds of fluid environments.

Much of the theory behind Rossby waves was developed in geophysical contexts---where one can assume that the thickness of the atmosphere or ocean in question is small compared to the size of the Earth---allowing for calculations to be done in two dimensions or in a quasi-2D system, such as the shallow-water system \citep[e.g.,][]{Pedlosky1987,2017aofd.book.....V}. The most common implementation of the shallow-water system uses an approximation in Cartesian coordinates called the $\beta$-plane and assumes the gravitational acceleration is constant across the entire domain. Additionally, these calculations tend to assume a stable, non-convective atmosphere. The Sun is neither a thin spherical shell of gas nor is the entirety of its interior convectively stable; so, any attempt to analyze the radial behavior of Rossby waves in a solar environment must first extend the present paradigm to more relevant stratifications. 

The helio- and astero-seismology communities have historically taken a very different approach to understand so-called ``$r$ modes" as they appear in stars. The field has traditionally relied on spheroidal oscillations (i.e., the $p$ modes and $g$ modes), which involve radial motions \citep[e.g.,][]{1983tsp..book.....C}. Without the presence of the Coriolis force, toroidal oscillations have zero frequency in the absence of damping. With the addition of rotation, toroidal modes appear in a form that resembles the Rossby waves noted by the geophysics community \citep{1949MSRSL...9....3L,1978MNRAS.182..423P}. Attempts to calculate the radial behavior of Rossby waves via this method immediately run into the problem that for low-frequency toroidal oscillations, there is coupling between the spheroidal and toroidal modes that results in an infinite series. To solve this problem, these calculations use asymptotic expansions with the rotation rate $\Omega$ of the star serving as the small parameter, to derive expressions for the radial behavior of these $r$ modes \citep{1981A&A....94..126P,1986SoPh..105....1W,2020A&A...637A..65D}.

The previous calculations of the radial variation of Rossby waves from both of these communities have prominently featured equations where the buoyancy frequency appears in terms of its reciprocal, $1/N^2$, and derivatives of its reciprocal. (The Ledoux discriminant $A$ is often used instead, which is related to the buoyancy frequency as $N^2 = gA/r$.) Because the square of the buoyancy frequency transitions from positive in the radiative interior to negative in the convection zone, this produces an internal singularity in the domain that must be handled when analyzing the entire solar interior. Calculations from the geophysics community often only treat stable atmospheres, which neatly avoids this problem \citep[e.g.,][]{Pedlosky1987, 2017aofd.book.....V}. In the stellar physics community, \citet{1981A&A....94..126P} do not attempt to treat realistic stratifications, opting to consider either fully convective or fully radiative stars, while \citet{2020A&A...637A..65D} focus specifically on polytropic atmospheres to model the convection zone only, and  \citet{2023A&A...671A..91A} focus on stable stratifications only.

Our most significant deviation from these previous calculations is to use the Lagrangian pressure fluctuation $\delta P$ as our working variable, rather than the Lagrangian displacement $\bm{\xi}$ \citep[e.g.,][]{1981A&A....94..126P,1986SoPh..105....1W,2020A&A...637A..65D}, latitudinal velocity $v_y$ \citep[e.g.,][]{2023A&A...671A..91A}, or the Eulerian pressure fluctuation $P_1$ \citep[e.g.,][]{Pedlosky1987}. The Lagrangian pressure fluctuation has been a convenient variable to work with in previous helioseismology calculations \citep[e.g.,][]{jcd2002, Gough1993} and often results in cleaner mathematics \citep[see][]{Hindman2022, Hindman2024}. We find that this method results in a greatly simplified vertical wave equation without internal singularities, even when the buoyancy frequency $N^2$ (or the Ledoux discriminant) changes sign. 

Rossby wave calculations often take the so-called ``traditional approximation", which ignores all Coriolis terms in the momentum equation that arise from the horizontal component of the rotation vector. While such an approximation is quite suitable for thin shells \citep[e.g.,][]{Eckart1960}, it is also often used by the asteroseismology community in regions with strong stratification, i.e., when $|\Omega/N| \ll 1$. Under such guidance, one would expect that the traditional approximation should be valid in the radiative interior and upper reaches of the convection zone of a low-mass star like the Sun\citep[e.g.,][]{1978A&A....70..597B}. This belief that the traditional approximation holds in such regions is derived from the study of gravito-inertial waves. To be careful, we elect to avoid the traditional approximation and in doing so we find that it is not valid in the Sun's radiative interior for Rossby waves.

In this work, we will use the standard $\beta$-plane approximation and take the limit of low-frequency waves by expanding the fluid variables in powers of $\omega/\Omega$ to calculate the vertical behavior in terms of the Lagrangian pressure fluctuation $\delta P$ (Section \ref{sec:equationset}). We then numerically calculate radial eigenfunctions and corresponding eigenfrequencies for a standard model of the Sun's radial stratification, Model S \citep{1996Sci...272.1286C}. We find two distinct families of modes, one that resides in the convection zone and the other in the radiative interior (Sections \ref{sec:eigenfunctions} and \ref{sec:eigenfrequencies}). Finally, we explore the implications of our calculations on the observability of the $r$ modes whose cavity lies within the radiative interior (Section \ref{sec:discussion}), as well on interpretations of the modes that have been previously observed.

\vfill\null


\section{Governing Equations} \label{sec:equationset}

We use the standard $\beta$-plane approximation with the origin of the local Cartesian coordinate system located at an arbitrary latitude $\theta$ and radius $R$. The coordinate unit vectors, $\unitv{x}$, $\unitv{y}$, and $\unitv{z}$, respectively point eastward, northward, and radially. The linearized, inviscid fluid equations are given by

\begin{align}
    & \rho_0 \pdv{u_x}{t} - f \left( \rho_0 u_y\right) + \tilde{f} (\rho_0 u_z) = -\pdv{P_1}{x}\;,
    \label{eq:xmom}\\
    & \rho_0 \pdv{u_y}{t} + f \left( \rho_0 u_x \right) = -\pdv{P_1}{y}\;,
    \label{eq:ymom}\\
    & \rho_0 \pdv{u_z}{t} - \tilde{f} (\rho_0 u_x) = -\pdv{P_1}{z} - g\rho_1 \;,
    \label{eq:zmom}\\
    & \pdv{\rho_1}{t} + (\bm{u} \cdot \bm{\nabla}) \rho_0 + \rho_0 \bm{\nabla} \cdot \bm{u} = 0\;, \label{eq:gencont}\\
    & \pdv{P_1}{t} + (\bm{u} \cdot \bm{\nabla}) P_0 + \gamma P_0 \bm{\nabla} \cdot \bm{u} = 0\;, \label{eq:genenergy}
\end{align}

\noindent where $\bm{u}=u_x\unitv{x}+u_y\unitv{y}+u_z\unitv{z}$ is the fluid velocity, $\rho_0 (z)$ and $P_0 (z)$ are the density and pressure of the hydrostatic background, and $\rho_1$ and $P_1$ are the Eulerian perturbations of the density and pressure. The gravitational acceleration $g$ is assumed constant, as is the adiabatic exponent $\gamma$. The Coriolis parameter is $f = 2 \Omega \sin \theta$, where $\Omega$ is the uniform rotation rate. The ``non-traditional" or ``horizontal" Coriolis parameter is given by $\tilde{f} = 2 \Omega \cos \theta$. As is typical in $\beta$-plane calculations, we will linearly expand the Coriolis parameter about a constant latitude $\theta_0$, such that $f = 2\Omega\sin\theta_0 + \beta y + \mathcal{O}\left(y^2/R^2\right)$ and $\beta = 2\Omega \cos\theta_0/R$.  In our final wave equation, we will only retain terms that are constant in the $y$ coordinate, i.e., $f \to f_0 = 2 \Omega \sin \theta_0$, $\partial_y f \to \beta = 2 \Omega \cos \theta_0 /R$, and $\tilde{f} \to \tilde{f}_0 = 2 \Omega \cos \theta_0$ .

We transform our primary variables from velocity to momentum density and introduce the Lagrangian pressure fluctuation, $\delta P$,

\begin{align}
    &\rho_0 \bvec{u} = \mathcal{U}\unitv{x} + \mathcal{V} \unitv{y} + \mathcal{W} \unitv{z} \;,\\
    & \pdv{}{t}\delta P \equiv \pdv{P_1}{t} + \bm{u} \cdot \bm{\nabla} P_0 = \pdv{P_1}{t} - g\mathcal{W}\;.
\end{align}

\noindent With these substitutions and after eliminating the vertical momentum density $\mathcal{W}$,

\begin{equation}
\mathcal{W} = \frac{i\omega}{g}\left(\delta P -P_1\right) \, ,
\end{equation}
our equation set becomes 

\begin{align}
    & -i \omega \mathcal{U} - f \mathcal{V} +  \frac{i \omega \tilde{f}}{g} \left(\delta P - P_1 \right) = -ik_x P_1 \;, \label{eq:momxtext} \\
    & -i \omega \mathcal{V} + f \mathcal {U} = -\pdv{P_1}{y}\;, \label{eq:momytext} \\
    & \frac{\omega^2}{g} \left( \delta P - P_1 \right) - \tilde{f} \mathcal{U} = -\pdv{P_1}{z} - \rho_1 g \label{eq:momztext} \\
    & -i \omega \rho_1 + i k_x \mathcal{U} + \pdv{\mathcal{V}}{y} + \frac{i \omega}{g} \pdv{}{z}\left( \delta P - P_1 \right) = 0 \;, \label{eq:conttext} \\
    \begin{split}
        & \frac{i\omega}{g} \left[\left(\frac{\partial}{\partial z} -\frac{1}{H_\rho}\right)(\delta P-P_1) + \frac{g}{c^2} \delta P\right] \\
        & \hspace{1.5in} = i k_x \mathcal{U} + \pdv{\mathcal{V}}{y} \;, \label{eq:energytext}
    \end{split}
\end{align}

\noindent where we have assumed longitudinal plane-wave solutions of the form $\sim \exp{i(k_xx - \omega t)}$, and we have written the vertical stratification in terms of the sound speed $c$, buoyancy frequency $N$, and density scale height $H_\rho$,

\begin{align}
    c^2 &= \gamma \frac{P_0}{\rho_0} \; , \\
    N^2 &= g\left(\frac{1}{H_{\rho}} - \frac{g}{c^2} \right)\;, \\
    H_{\rho}^{-1} &= -\frac{1}{\rho_0}\dv{\rho_0}{z}\;.
\end{align}

In order to correctly capture the quasi-geostrophy indicative of Rossby waves, we must consider a low-frequency approximation to these equations (as opposed to a slow-rotation expansion). We use $\beta/(k_x \Omega)$ as a small parameter and assume $\omega \sim \beta/k_x \ll \Omega$ \citep[see:][]{Pedlosky1987}. Equations~(\ref{eq:momxtext})--(\ref{eq:energytext}) make it clear that to leading order in the wave frequency, the horizontal fluid motions are geostrophic. However, in order to achieve Rossby wave propagation, we must break the pure geostrophic balance by keeping terms up to first order in the frequency.  Hence, at the moment we keep all terms in the horizontal momentum equations, continuity equation and energy equation. In particular, we note that the non-traditional Coriolis term in the $\hat{x}$-momentum equation \eqref{eq:momxtext} is the same order as the inertial term, and hence for self-consistency, must be retained. In the vertical momentum equation, it is immediately clear that the inertial term in the vertical equation \eqref{eq:momztext} is second order and can be dismissed and that the non-traditional Coriolis term is zero-order and must be kept. This approximation where only the inertial term in the vertical equation is neglected is known as the ``quasi-hydrostatic approximation" and holds as long as $\tilde{f}$ can be treated as spatially constant \citep{1995QJRMS.121..399W,1997JGR...102.5733M,2005QJRMS.131.2081W,2008RvGeo..46.2004G}.

Using the vertical momentum equation to eliminate $\rho_1$ in the continuity and energy equations, then using the continuity equation to rewrite the divergence in the energy equation, we have a reduced equation set of

\begin{align}
     & -i \omega \mathcal{U} - f \mathcal{V} + \frac{i \omega}{g} \tilde{f}  \left(\delta P - P_1 \right) = -i k_x P_1 \;,
    \label{eq:xmom4}\\
    & -i \omega \mathcal{V} + f \mathcal{U} = -\pdv{P_1}{y}\;,
    \label{eq:ymom4}\\
    & ik_x \mathcal{U} + \pdv{\mathcal{V}}{y} = - \frac{i \omega}{g} \left(\pdv{\delta P}{z} - \tilde{f} \mathcal{U} \right)  +\mathcal{O}\left(\omega^2/\Omega^2\right) \label{eq:cont4} \\
    & \left(\pdv{}{z} + \frac{1}{H_{\rho}} \right) P_1 = \frac{N^2}{g} \delta P + \tilde{f} \mathcal{U} +\mathcal{O}\left(\omega^2/\Omega^2\right) \label{eq:energy4}.
\end{align}

Since Rossby waves are a vorticity wave, the cleanest way forward is to derive an equation for the vertical component of the vorticity and reduce that equation to a PDE for a single dependent variable. By taking the curl and horizontal divergence of Equations~\eqref{eq:xmom4} and \eqref{eq:ymom4}, one obtains

\begin{align}
    -i\omega \mathcal{Z} &= -f_0 \Delta - \beta \mathcal{V} + \frac{i \omega}{g} \tilde{f}_0 \pdv{}{y} \left( \delta P - P_1 \right)  \;, \label{eq:vorticity}\\
    -i\omega \Delta &= f_0 \mathcal{Z} - \left(\frac{\partial^2}{\partial y^2}-k_x^2\right)P_1 + \frac{ \omega k_x}{g} \tilde{f}_0 \left( \delta P - P_1 \right) \;,  \label{eq:dilation}
\end{align}

\noindent where the vorticity $\mathcal{Z} = ik_x\mathcal{V} - \partial_y \mathcal{U}$ and horizontal dilation $\Delta = ik_x\mathcal{U} + \partial_y \mathcal{V}$ are density weighted.

If we specify that the vorticity and horizontal velocity components are all order one, Equation~\eqref{eq:cont4} then dictates that the horizontal divergence is small by a factor of $\omega/\Omega$ and so too are the inertial terms in the momentum equations \eqref{eq:xmom4} and \eqref{eq:ymom4}.  Therefore, to lowest order in the wave frequency the motions are geostrophic,

\begin{align}
    &  \mathcal{V} \approx \frac{ik_x}{f_0} P_1 + \mathcal{O}\left(\omega/\Omega\right) \; , \label{eq:geo_xmom} \\
    & \mathcal {U} \approx -\frac{1}{f_0} \pdv{P_1}{y} + \mathcal{O}\left(\omega/\Omega\right) \;, \label{eq:geo_ymom}
\end{align}

\noindent and the dilation equation \eqref{eq:dilation} becomes to leading order

\begin{equation}
    \mathcal{Z} \approx \frac{1}{f_0} \left(\frac{\partial^2}{\partial y^2}-k_x^2\right)P_1 + \mathcal{O}\left(\omega/\Omega\right)\;.  \label{eq:geo_dilation}
\end{equation}

Using \cref{eq:cont4,eq:geo_xmom,eq:geo_dilation} to eliminate $\Delta$, $\mathcal{V}$, and $\mathcal{Z}$ from the vorticity equation \eqref{eq:vorticity}, we obtain two equations in terms of the Lagrangian pressure fluctuation and the Eulerian pressure fluctuation. To leading order in the frequency, these equations are

\begin{align}
    \frac{g}{f_0^2} \left(k_x^2 - \frac{\partial^2}{\partial y^2} - \frac{\beta k_x}{\omega} \right) P_1 &= \left( \pdv{}{z} + \frac{\tilde{f}}{f_0} \pdv{}{y} \right) \delta P \; ,\\
    \left( \pdv{}{z} + \frac{\tilde{f}}{f_0} \pdv{}{y} + \frac{1}{H_{\rho}} \right) P_1 &= \frac{N^2}{g} \delta P \; .
\end{align}

\noindent The first-order derivatives in $y$ are a direct result of keeping the horizontal Coriolis term $\tilde{f}$. Notably, the specific combination of $z$ and $y$ derivatives in both equations is actually an axial derivative along the rotation axis. This indicates plane wave solutions that travel in longitude and cylindrical radius are appropriate, i.e., solutions should have the form,
\begin{equation}
   F=\exp\left[i (k_x x + k_y y - \omega t)\right] \, \exp\left(-i k_y \cot\theta_0\,z\right) \; \hat{F}(z) \,,
\end{equation}
\noindent where $F$ is any wave variable and $\hat{F}(z)$ is the vertical component of its eigenfunction beyond the plane-wave hypothesis. (Similar plane-wave solutions are used in \citet{2025JFM..1007A..61P}.) Upon refining our initial plane-wave ansatz, we then eliminate $P_1$ and obtain an ODE in terms of only the Lagrangian pressure fluctuation $\delta\hat{P}$:

\begin{equation} \label{eq:ode}
\begin{split}
    \frac{1}{f_0^2}& \left(k_x^2 + k_y^2 + \frac{\beta k_x}{\omega} \right) \delta\hat{P} = \\
    &\hspace{0.5in} \frac{1}{N^2} \left( \pdv{}{z} + \frac{1}{H_{\rho}} \right) \pdv{\,\delta\hat{P}}{z} = -\frac{\delta\hat{P}}{g h}
\end{split}
\end{equation}

\noindent where $1/gh$ is the separation constant. Equation \eqref{eq:ode} presents a tractable vertical equation, which we will unpack in the next section, and a familiar horizontal equation identical to the equation for Rossby waves in a shallow-water system. Hence, our eigenvalue $h$ serves as the ``effective depth" of the fluid layer. 

The momentum densities and the Eulerian pressure fluctuation can be expressed in terms of $\delta\hat{P}$ as follows:

\begin{align}
   \hat{\mathcal{U}} &= \frac{i k_y h}{f_0} \pdv{\delta\hat{P}}{z}  \label{eq:u}\\
    \hat{\mathcal{V}} &= -\frac{i k_x h}{f_0} \left( \pdv{\delta\hat{P}}{z} \right)  \label{eq:v}\\
    \hat{\mathcal{W}} &= \frac{i \omega}{g} \left(1 + h \pdv{}{z} \right) \delta\hat{P} \label{eq:w}\\
    \hat{P}_1 &= -h \pdv{\delta\hat{P}}{z}  \label{eq:p1}
\end{align}

\noindent Each of these variables is dependent upon the vertical derivative of $\delta\hat{P}$.

\section{Wave Cavities and Vertical Eigenfunctions} \label{sec:eigenfunctions}

\propdiagram

{The vertical wave equation is second-order in $z$}

\begin{align} \label{eq:verteq}
    \frac{d^2 \, \delta\hat{P}}{d z^2} + \frac{1}{H_{\rho}} \dv{\,\delta\hat{P}}{z} + \frac{N^2}{gh} \delta\hat{P} = 0 \;.
\end{align}

\noindent and can be rewritten as a Sturm-Liouville equation,

\begin{align} \label{eq:sturm}
    \dv{}{z} \left(\frac{1}{\rho_0} \dv{\,\delta\hat{P}}{z} \right) = -\frac{1}{\rho_0}\frac{N^2}{g h} \delta\hat{P},
\end{align}

\noindent where $h$ is the eigenvalue and $N^2/(\rho_0 g)$ is the weight function. Because $N^2$ is positive in the radiative interior and negative in the convection zone, the weight function changes sign. For a system such as this with an indeterminate weight function, there are two families of real eigenvalues, with each family possessing a countable infinity of distinct eigenvalues \citep[e.g.,][]{Ince1956}. One family has positive eigenvalues that stretch unbounded to $+\infty$ and eigenfunctions oscillating in the radiative interior where the weight function is positive. The other family has negative eigenvalues (reaching towards $-\infty$) and eigenfunctions oscillating in the convection zone where the weight function is negative. We will label the radial order of the eigenfunctions corresponding to the positive and negative families of eigenvalues $n_{ri}$ and $n_{cz}$, respectively.

To explore the differences in the two wave cavities, we perform a variable transform $\delta\hat{P} = \sqrt{\rho_0} \, \psi$ to convert the ODE into standard form,

\begin{equation} \label{eq:vert}
    \frac{d^2 \psi}{d z^2} + k_z^2(z) \, \psi = 0 \;,
\end{equation}

\noindent where $k_z$ is a local vertical wavenumber,

\begin{equation} \label{eq:local_dispersion}
    k_z^2(z) = N^2 \left(\frac{1}{gh} - \frac{\omega_c^2}{c^2 N^2} \right) \;,
\end{equation}

\noindent and $\omega_c$ is the acoustic cut-off frequency,

\begin{equation}
    \omega_c^2 \equiv \frac{c^2}{4H_\rho^2} \left(1-2 \frac{dH_\rho}{dz}\right) \;.
\end{equation}

\noindent Figure \ref{fig:propdiagram} shows the propagation diagram for a standard solar model, Model S from  \cite{JCD1996}. The thick black curve corresponds to the vertical profile of $\omega_c^2/(c^2 N^2)$. We expect vertical propagation wherever $k_z^2$ is positive. Hence, in the radiative interior where $N^2 > 0$, for propagation $1/gh$ must exceed $\omega_c^2/(c^2 N^2)$, i.e., the separation constant must lie above the black line in Figure \ref{fig:propdiagram}. Conversely in the convection zone, where $N^2 < 0$, the quantity $1/gh$ must be more negative than $\omega_c^2/(c^2 N^2)$ and thus lies below the black line. We note two clear wave cavities as expected, marked in orange and purple, respectively.

Assuming boundary conditions of $\delta \hat{P} = 0$ at the top of the solar model ($r = 1.05 \, R_\odot$) and $\partial_z \delta \hat{P} = 0$ at solar center ($r = 0$), we use the shooting method to numerically solve for our eigenvalues $h_n$ for both cavities in Model S, which are plotted as $1/gh_n$ with blue lines in the radiative interior and yellow lines in the convection zone. Again as expected, the two families of eigenvalues are unbounded.

The first eleven eigenvalues for both families of modes are presented in Table \ref{tab:eigenvalues} (and shown in Figure \ref{fig:propdiagram}). For a given value of $g$, we calculate the ``effective depth" of these modes $h$. For the radiative interior modes, we take $g = 5.18 \times 10^{4}$ cm s$^{-2}$, a value appropriate for the base of the convection zone.  For the convection zone modes, we utilize the photospheric value of the gravitational acceleration, $g = 2.74 \times 10^{4}$ cm s$^{-2}$. The radiative interior modes have an effective depth that decreases with increasing radial order $n_{ri}$. Due to their unstable stratification, the convection zone modes have an effective depth of value $-h$, which decreases with increasing $n_{cz}$.

Next, we look at the vertical structure of the modes absent the plane wave component $\exp(-i k_y \cot \theta_0 z$), which has no impact on the eigenvalues. Figure \ref{fig:eigenfunctions} shows the three lowest-order radial eigenfunctions for model S in Lagrangian pressure $\delta\hat{P}$ (\textit{left}), reduced Lagrangian pressure  $\delta\hat{P}/\rho_0$ (\textit{middle}), and vertical vorticity $\hat{\zeta}_z = \hat{\mathcal{Z}}/\rho_0$ (\textit{right}) for both radiative interior modes (\textit{top}) and convection zone modes (\textit{bottom}). All fluid variables for the radiative-interior are vertically propagating throughout the radiative interior and evanescent in the convection zone, while the lowest order convection-zone modes are confined just below the photosphere. Notably, the vorticity eigenfunctions for the radiative interior have exceedingly long evanescence lengths in the convection zone and maintain a large amplitude all the way through the convection zone, even at the photosphere. We defer further discussion of this result to Section \ref{sec:discussion}. 

The plane wave component introduces a dependence on both  the latitude $\theta_0$ of our $\beta$-plane and the latitudinal wavenumber $k_y$. This dependence is naturally weak for the sectoral modes in both the radiative interior and convection zone cavities (because $k_y R$ is small).  Further, since the convection zone modes are confined to a narrow vertical cavity, the vertical variation over the cavity itself is additionally small. As such, the vertical behavior for the convection zone modes and the sectoral modes will look as pictured in Figure \ref{fig:eigenfunctions}. For tesseral radiative interior modes, this additional plane wave will not impact the location of the wave cavities but will make the plotted envelopes oscillatory.

\eigenfunctions

\eigenvalues

\vfill\null

\section{Eigenfrequencies} \label{sec:eigenfrequencies}

The separated horizontal equation derived from Equation~(\ref{eq:ode}) forms the global dispersion relation once the vertical eigenvalues $h_n$ have been generated by the vertical equation~(\ref{eq:verteq}),

\begin{align}
   & \frac{1}{f_0^2} \left(\frac{\beta k_x}{\omega} + k_x^2 + k_y^2\right) = -\frac{1}{gh_n} \;.
\end{align}

\noindent Solving for the wave frequency produces the familiar form,

\begin{align}
    \omega_n(k_x,k_y) = -\frac{k_x \beta}{k_x^2 + k_y^2 + f_0^2/gh_n}. \label{eq:dispersion}
\end{align}

Noting that the dispersion relation in the 2-D case is 

\begin{equation}
    \omega_{2D} = -\frac{k_x\beta}{k_x^2 + k_y^2},
\end{equation}

\noindent we see that eigenvalues $h_n$ for the two families of solution will shift the frequencies in opposite directions. Figure \ref{fig:dispersion} shows the fractional frequency shift with respect to the two-dimensional case $(\omega_n - \omega_{2D})/\omega_{2D}$, assuming a latitude of $\theta = 10$ degrees, for the first seven radial modes $n = 0-6$ in the (a) radiative interior and (b) convection zone cavities. We calculated $h_n$ numerically by solving the vertical equation for Model S. The radiative interior modes, with positive values of $h_n$ will shift the frequency to be more positive. The convection zone modes, with negative values of $h_n$, will shift the frequencies to be more negative. This impact of stable versus unstable stratification on Rossby wave frequencies has been noted before \citep[e.g.,][]{1968RSPTA.262..511L,2017aofd.book.....V,2020A&A...637A..65D,1986SoPh..105....1W}. 

Though the precise frequencies are highly dependent upon the latitude of the $\beta$-plane, the fractional frequency shift provides a general qualitative sense of what to expect. The fractional frequency shift for the radiative-interior case is so small as to be unobservable, with the largest shifts having magnitudes of one part in ten thousand. The convection zone modes exhibit significantly larger shifts, particularly for low azimuthal order $m$. While $m = 1$ and $m = 2$ are strongly polluted by the Cartesian approximation and probably should not be trusted, we still see shifts of up to 10\% for modes of higher azimuthal order ($m \ge 3$). 

\section{Discussion} \label{sec:discussion}

In this work, we used the Lagrangian pressure fluctuation $\delta P$ to derive a very simple vertical equation to describe Rossby waves in a generally stratified atmosphere. We then solved this equation for Model S,  discovered two wave cavities with two families of Rossby waves, calculated the vertical eigenfunctions for each family, and predicted the corresponding mode frequencies.

\vfill\null

\subsection{Validity of the traditional approximation}

We elected not to take the ``traditional approximation", where the horizontal component of the rotation vector of the Coriolis force $\tilde{f}$ is neglected \citep[e.g.,][]{Pedlosky1987,2008RvGeo..46.2004G}. Since the traditional approximation is generally valid for thin fluid layers, it is commonly employed in geophysics \citep[e.g.,][]{Eckart1960}. In asteroseismic applications, the traditional approximation is often used whenever $|\Omega/N|$ is deemed small \citep[e.g.,][]{1978A&A....70..597B,1988BCrAO..80..151G} as happens in both the Sun's radiative interior and at the top of its convection zone. However, as we have seen, including the nontraditional terms results in the presence of an additional plane wave component $\exp\left(-i k_y \cot(\theta) z\right)$. From this we can easily deduce that the nontraditional Coriolis terms can be ignored only if the latitudinal wavelength is much longer than the vertical extent of the wave cavity. This is a far more stringent restriction than is commonly assumed. Its probably safe to ignore the nontraditional terms for sectoral modes which have very long latitudinal wavelengths on the order of several stellar radii. Further, the convection zone modes have very narrow vertical cavities confined in the upper portion of the convection zone such that the nontraditional terms are likely to be unimportant. However, tesseral modes of the radiative interior probably violate the traditional approximation.
The perhaps erroneous belief that the traditional approximation holds in the radiative interior derives from the study of gravito-inertial waves which have very short vertical wavelengths in highly stratified atmospheres.  On the other hand, tesseral Rossby modes of low radial order have vertical lengths comparable to the depth of the interior of the stellar model (see Figure~\ref{fig:eigenfunctions}), and we should not expect them to behave in the same way.

\dispersion

\subsection{$\delta P$ is a convenient variable}

Equation \eqref{eq:sturm} is a Sturm-Liouville problem with real and nonsingular coefficients that arises from the choice of the Lagrangian pressure fluctuation $\delta P$ as the primary variable. Importantly, there are no internal singularities, even when $N^2$ passes through zero as it transitions from the stable radiative interior to the unstable convection zone. This differs from previous works where the buoyancy frequency (or equivalently, the Ledoux descriminant) appears in the denominator of the PDE coefficients. Other studies that employed the $\beta$-plane approximation (as we have done here) have instead used Eulerian pressure \citep[][]{Pedlosky1987} or latitudinal velocity \citep[][]{2023A&A...671A..91A} as their primary variables. From \cref{eq:u,eq:v,eq:w,eq:p1} above, we see that each velocity component and the Eulerian pressure fluctuation all rely on the vertical derivative of the Lagrangian pressure fluctuation, which explains the appearance of internal singularities in these other calculations. By taking the vertical derivative of our vertical equation \eqref{eq:sturm} and substituting equation \eqref{eq:v} one can generate an ODE with derivatives of $1/N^2$: 

\begin{equation} \label{eq:ode_vy}
    \frac{d}{dz}\left(\frac{1}{N^2}\frac{du_y}{dz}\right) - \frac{1}{H_{\rho}N^2} \dv{u_y}{z} +\frac{u_y}{gh} = 0,
\end{equation}

\noindent which is equivalent to Equation (15) in \cite{2023A&A...671A..91A}. Equations \cref{eq:u,eq:v,eq:w,eq:p1} show that the singularity is removable: none of the wave variables themselves are singular, and there are not additional modes associated with the singularity (such as an $f$-mode like boundary mode at the interface between positive and negative $N^2$, or critical latitude modes arising from a singularity when the wave's phase speed equals the local rotational speed.) While such internal singularities may not be removable in problems that do not take the $\beta$-plane approximation, the use of the Lagrangian pressure fluctuation certainly simplifies the mathematics in this study.

One advantage of the Sturm-Liouville form of the vertical wave equation~\eqref{eq:sturm} is that standard perturbation analysis techniques can be applied in a straightforward manner to derive sensitivity kernels for use in helioseismology. Direct inspection of Equation~\eqref{eq:sturm} reveals that the eigenvalues (and hence the frequencies) are sensitive to the mass density and the buoyancy frequency. While the Sun's $p$ modes have long been used to robustly deduce the mass density, the buoyancy frequency within the convection zone is poorly constrained. Hence, if the frequencies of multiple radial overtones could be measured, the Sun's Rossby waves could in principle be used to measure the radial variation in the superadiabaticity within the convection zone. Even if we are never able to resolve individual radial overtones to measure their frequencies, the collective properties of the spectrum of modes might contain information about gross spatial averages of the superadiabaticity.

While this paper demonstrates that our choice of variable can greatly simplify the vertical equation, it has its limitations. Because we made the Cartesian $\beta$-plane approximation, we do not see the effects of curvature, which are undoubtedly important.

\subsection{Easily identifiable wave cavities}

Previous work \citep{1981A&A....94..126P, 1986SoPh..105....1W,1988BCrAO..80..151G,2020A&A...637A..65D} has recognized that there must exist two distinct wave cavities, one in the convection zone and another in the radiative interior.  However, by using the Lagrangian pressure fluctuation resulting in Equation \eqref{eq:vert}, it is exceedingly simple to derive a local dispersion relation \eqref{eq:local_dispersion} that enables quick identification of the location and spatial extent of these cavities. It is also worth noting that the Sun's convection zone ends just below the photosphere and the atmosphere above the photosphere is a region of convective stability, $N^2>0$. This suggests that a third cavity of Rossby waves could potentially exist within the Sun's upper atmosphere. We have not identified any modes of such a third cavity; but, to ascertain whether this region successfully traps waves in a vertical waveguide would require extending the propagation diagram throughout the Sun's temperature minimum into the chromosphere.  Model S only reaches a short distance above the photosphere and lacks a realistic temperature profile in the upper atmosphere.

\subsection{Comparison to radial observations}

The few attempts to observationally tease out the radial eigenfunctions for the solar Rossby modes disagree on the radial behavior. \cite{Proxauf2020} found something similar to $r^m$ in radial vorticity down to 8 Mm, followed by a decrease in power inconsistent with a power-law; \cite{Hathaway2021} find roughly constant amplitude down to 37 Mm; and \cite{2024ApJ...967...46M} see an increase in power down to 55 Mm, followed by a decrease in power until 118 Mm. 

\subsubsection{Radiative interior modes}
As seen in Figure \ref{fig:eigenfunctions}(c), the vertical vorticity eigenfunctions for the radiative interior modes are roughly constant throughout the convection zone, which seems to align with the findings of \citet{Hathaway2021}. In both Model S and the Sun, $N^2$ is extremely small in the convection zone, with $|N^2 H_{\rho}/g| < 10^{-5}$. Thus, Equation \eqref{eq:sturm} dictates that $\partial_z(\delta P/\rho_0)$ is a constant which implies through Equations \eqref{eq:u} and \eqref{eq:v} that the horizontal velocity components are also vertically constant.  This is precisely the behavior that we see in Figure \ref{fig:eigenfunctions}(c).

We should not immediately dismiss the possibility of observing the radiative interior modes; their velocity and vorticity eigenfunctions are essentially constant across the convection zone. Of course, we have ignored sphericity, and in a proper spherical geometry there are likely to be curvature terms that introduce power law behavior in radius. Additionally, we have ignored radial differential rotation, which may prevent such modes from reaching the surface via the formation of critical layers . Either way, the observation of wave modes living in the radiative interior is a possibility that we should take seriously. 

\subsubsection{Convection zone modes}

Previous theoretical work suggests that the eigenfunctions for modes of the convection zone cavity should vary with a dependence of $r^m$, where $m$ is the azimuthal order of the wave \citep{1981A&A....94..126P,2020A&A...637A..65D}. We do not see such behavior, and more stringently, our vertical wave equation \eqref{eq:sturm} lacks all dependence on the horizontal wavenumbers. These previous calculations were perturbative, expanding each variable in powers of angular frequency $\Omega$ to obtain radial eigenfunctions for a polytropic atmosphere. 

The eigenfunctions for the convection-zone cavity presented here are qualitatively similar to those presented by \cite{2024ApJ...967...46M}, though our eigenfunctions are confined to a much narrower region of about 4 Mm just below the photosphere. Given the possibility of radiative interior mode detection, it is possible that attempts to determine the radial behavior of the observed modes are seeing an entanglement of both the convection zone and radiative interior modes.  

\resolution

\subsection{Mode misidentification}

As we can see from Figure \ref{fig:dispersion}, the frequency spacing between radial overtones for the radiative-interior modes of Model S are negligible, while the spacing between the convection zone modes is on the order of nanohertz. Notably, the impact of an unstable stratification on the expected frequency is to shift the eigenfrequencies more negatively compared to the 2-D case. As the radial order increases, these frequencies continue to shift in this direction. 

There has been much discussion in the community about the set of modes originally identified by Bekki in numerical simulation as ``equatorial Rossby waves with one radial node" \citep{2022AA...662A..16B,2022AA...666A.135B, Waidele2023}. Because we (and others) find the direction of the frequency shift of the radial overtones of the convection zone modes to be more negative, this particular mode identification is contrary to our findings. This mode has already been noted to have a significant non-toroidal component, and this work simply agrees that this mode is something distinct from the classical toroidal Rossby waves that have been observed.

\subsection{Distinguishing between mode families}

Figure \ref{fig:resolution} displays artificial spectra for the $\ell = m = 3$ modes at two different frequency resolutions (a) 0.03 nHz, corresponding to 1000 years of data and (b) 3 nHz, corresponding to 10 years of data. The line profiles display the power contributed by the first eleven radial orders for both the radiative interior modes and the convection zone modes, with the $n_{ri} = 0$ frequency marked by the dashed orange line and the $n_{cz} = 0$ frequency marked by the dashed purple line. Each mode is assumed to possess a Lorenzian power profile, and the mode power is added incoherently. Each mode profile has the same full linewidth of 0.1 nHz and the prescribed amplitude of a mode falls off like $1/n^2$. 

The modes of the convection zone cavity ($n_{cz}$) appear to the left of the purple line, whereas all eleven of the modes of the radiative interior ($n_{ri}$) blend into a single peak at the location of the orange line, as the frequency spacing between modes of adjacent vertical order is far less than the linewidth. On the other hand, the frequency separations between vertical overtones of the convection zone modes ($n_{cz}$) are greater than the linewidth and are thus well-separated, with five or six radial overtones visible before falling below the arbitrary noise floor. With the more realistic resolution in the right panel, we can see that none of these modes are well-resolved and clump together into one large asymmetric peak. Observations of solar $r$-modes have power spread over several nHz, running from the expected 2-D value to more negative frequencies at low $m$, where one would expect stratification effects to dominate over differential rotation effects \citep[e.g.,][]{Loptien2018,2019A&A...626A...3L,Waidele2023,Hanson2024}. We speculate that we could be seeing a combination of the radiative interior modes and convection zone modes, potentially of multiple radial orders.

\begin{acknowledgements}
C. Blume was supported in this work by the Future Investigators in NASA Earth and Space Sciences Technology (FINESST) award 80NSSC23K1624. Hindman was supported through NASA grants 80NSSC24K0125, 80NSSC24K0271, and 80NSSC24K0927.
\end{acknowledgements}

\bibliography{Bibliography}

\end{document}

%% file: Figures.tex
\def\propdiagram{
\begin{figure}
    \centering
    \includegraphics[scale=0.5]{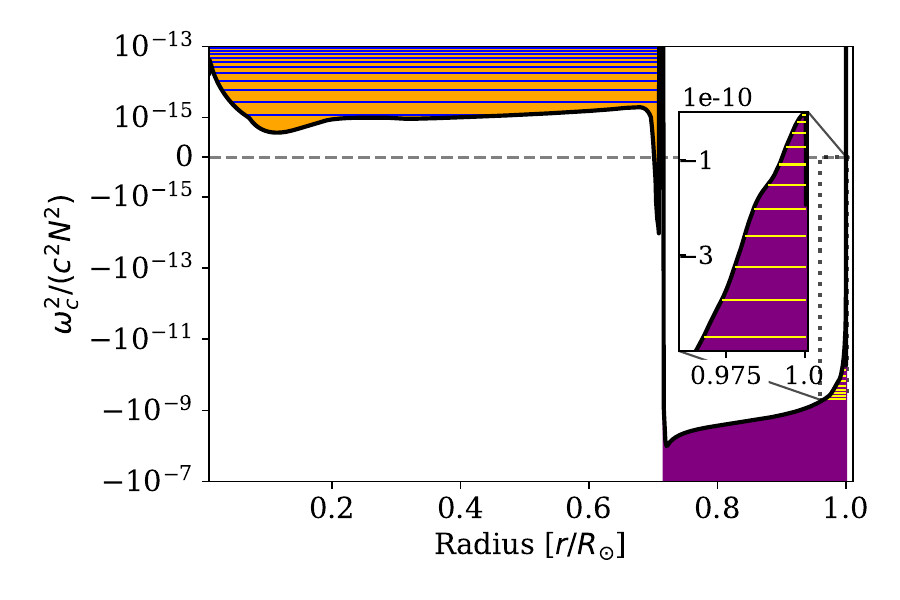}
    \caption{\normalsize \textbf{Propagation diagram for Model S}---The orange (purple) region denotes propagation for Rossby waves in the radiative interior (convection zone). The blue (yellow) lines mark the values of the separation constant for low radial order modes in each region respectively. The inset shows the very top of the convection zone, with $y$-axis values ranging from $-5 \times 10^{-10}$ to $0$.}
    \label{fig:propdiagram}
\end{figure}
}

\def\eigenfunctions{
\begin{figure*}
    \centering
    \includegraphics[scale=0.5]{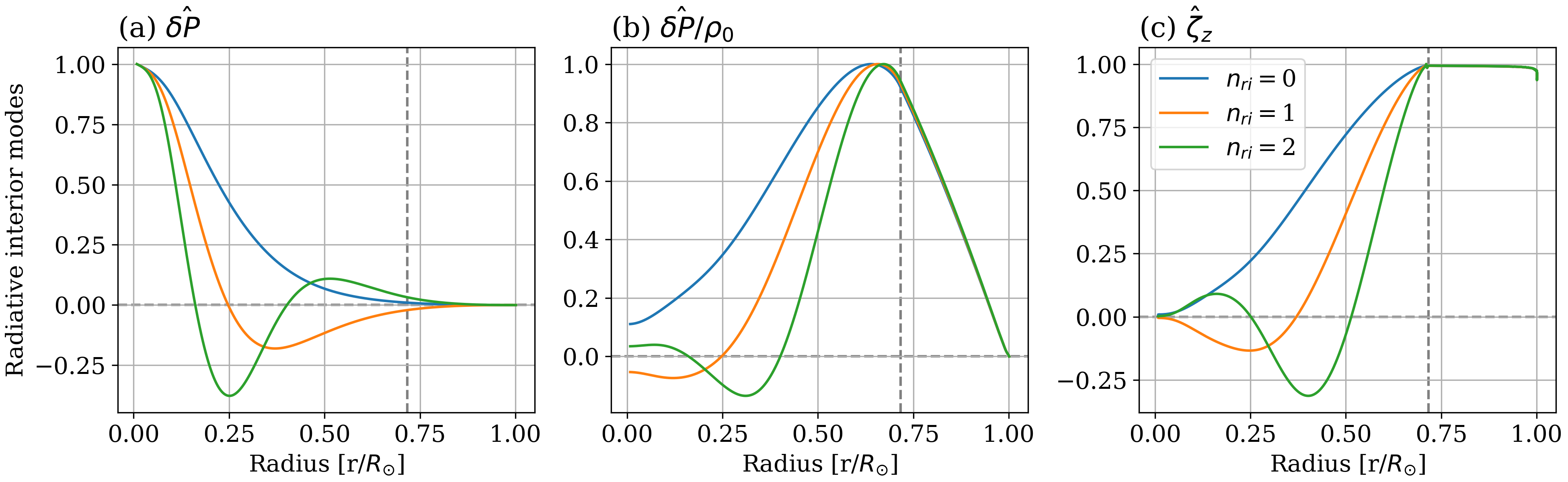}
    \includegraphics[scale=0.5]{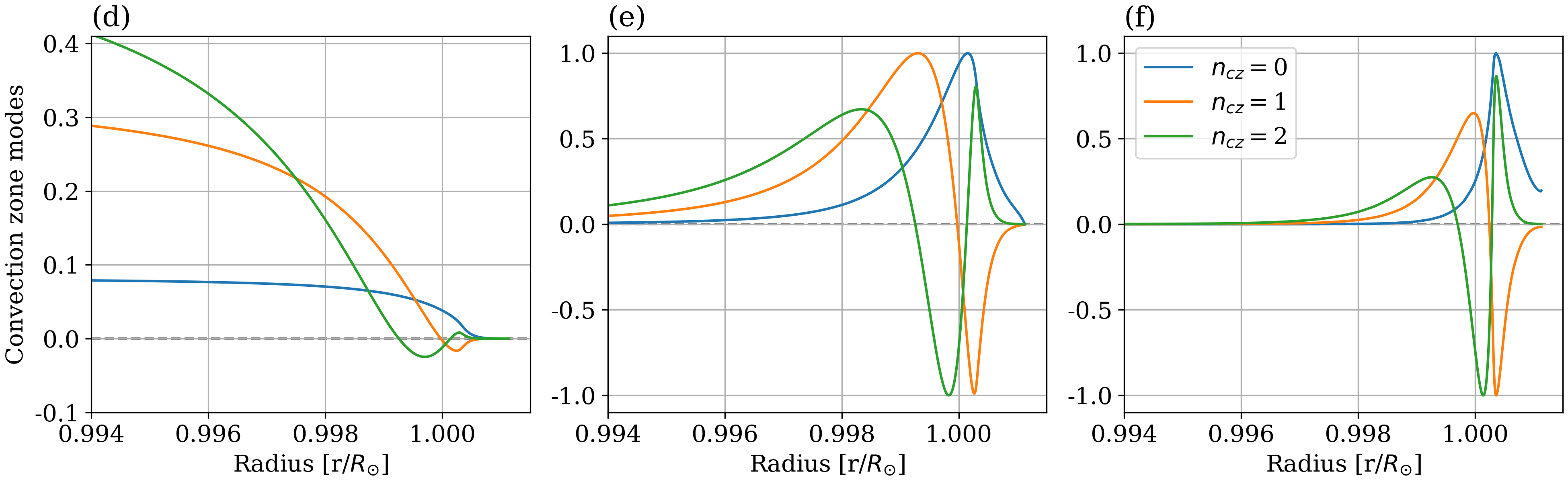}
    \caption{\normalsize  \textbf{Radial eigenfunctions for Rossby waves}---Radial eigenfunctions for the radiative interior modes (\textit{top}) and convection zone modes (\textit{bottom}) in Lagrangian pressure fluctuation $\delta \hat{P}$ (\textit{left column}), reduced pressure $\delta\hat{P}/\rho_0$ (\textit{middle column}) and radial vorticity $\hat{\zeta}_z$(\textit{right column}). The radiative interior mode eigenfunctions are distributed across the region and are roughly constant in the convection zone. The convection zone mode eigenfunctions are confined near the surface and evanescent everywhere else.}
     \label{fig:eigenfunctions}
\end{figure*}
}

\def\dispersion{
\begin{figure*}
    \centering
    \includegraphics[scale=0.55]{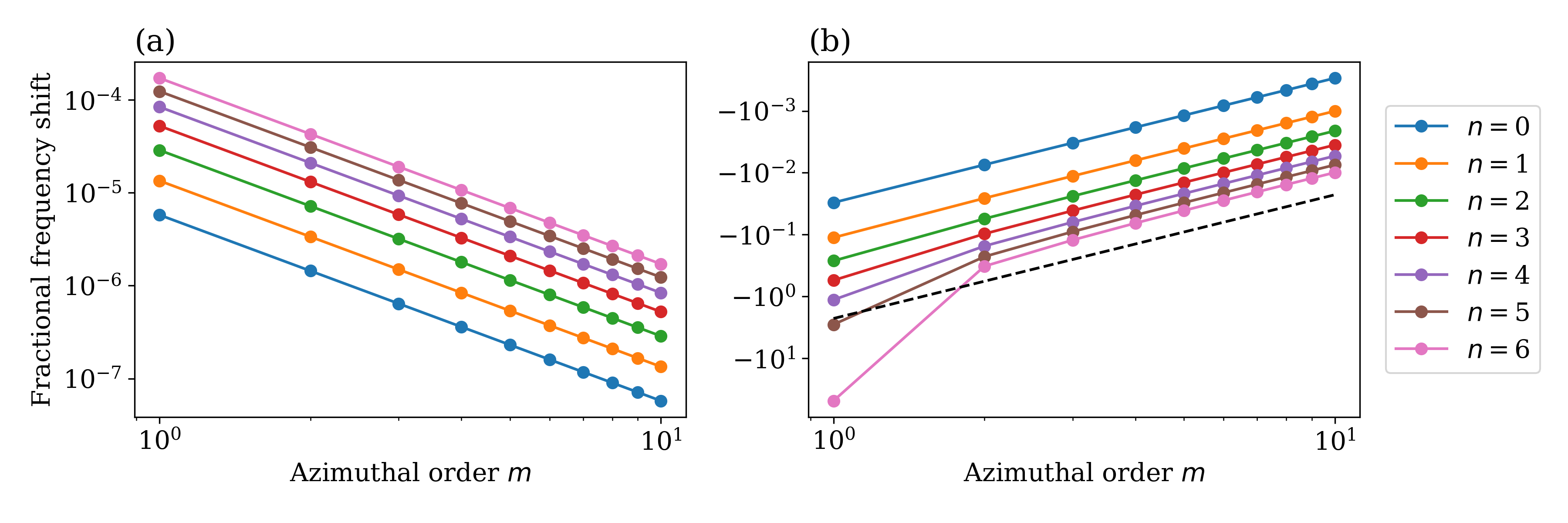}
    \caption{\normalsize  \textbf{Fractional frequency shift}---The fractional frequency shift $(\omega_n - \omega_{2D})/\omega_{2D}$ with respect to the two-dimensional dispersion relation for modes of the (a) radiative interior and (b) convection zone. The figure was generated for a low-latitude $\beta$-plane located at $\theta$ = 10 degrees. The black dashed line represents the asymptotic behavior for large $m$. The frequency shift for the radiative interior modes is so small as to be undetectable. On the other hand, the convection zone modes can have quite a large frequency difference in comparison. }
    \label{fig:dispersion}
\end{figure*}
}

\def\resolution{
\begin{figure*}
    \centering    \includegraphics[scale=0.5]{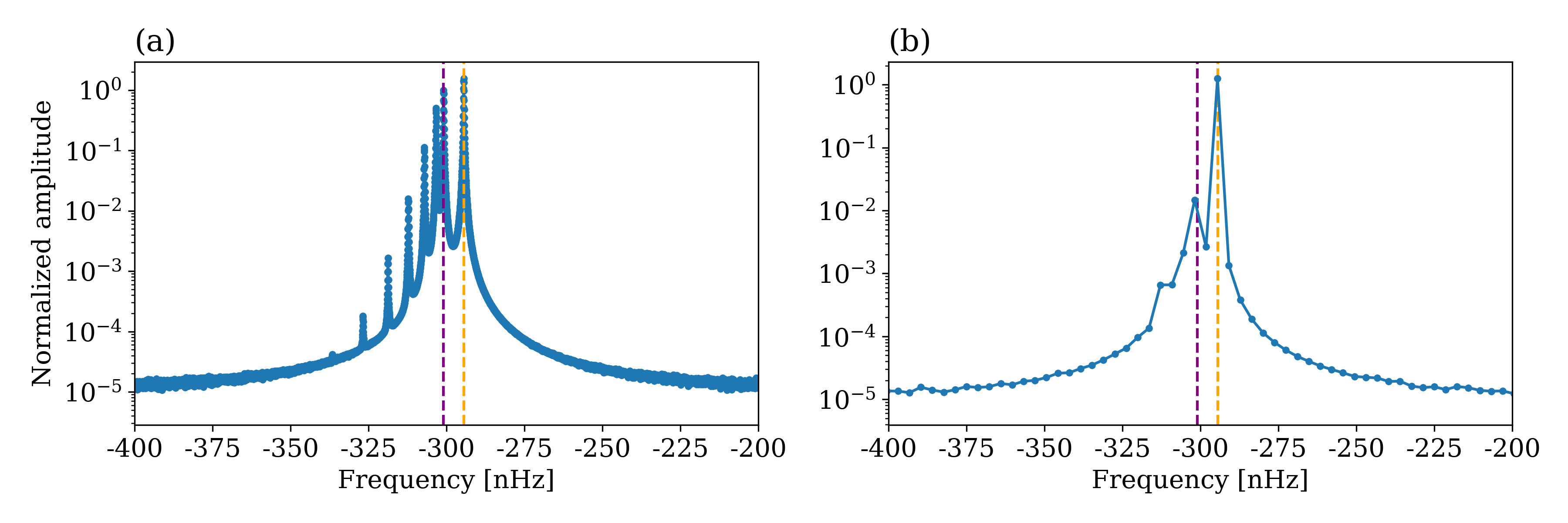}
    \caption{\normalsize \textbf{Artificial spectra}---Artificial spectra calculated for the $\ell = m = 3$ mode with different frequency resolution (a) 0.03 nHz (observations spanning 1000 yrs) and (b) 3 nHz (observations spanning 10 yrs). Each panel displays the first 11 modes of increasing radial order for both the radiative interior and convection zone families. Each mode peak has been given an arbitrary natural linewidth of 0.1 nHz and an amplitude that diminishes with radial order like 1/$n^n$. The zeroth order mode for the radiative interior (convection zone) family is marked with an orange (purple) dashed line. The unrealistic frequency resolution shown in the left panel resolves the individual mode peaks, while with the more realistic resolution on the right, the modes are unresolved and blend into a single peak with power fluctuations that could be confused with realization noise.}
    \label{fig:resolution}
\end{figure*}
}

\def\eigenvalues{
\begin{table*}
\centering
\textbf{Table~\ref{tab:eigenvalues}\\}
{Rossby Wave Effective Depths} 

\begin{tabular}{c|c|c||c|c|c}
\hline \hline

\multicolumn{3}{c|}{\textbf{Radiative Interior Modes}}  & \multicolumn{3}{|c}{\textbf{Convection Zone Modes}} \\
\hline

$n_{ri}$ & $1/gh_n$ & $h_n/R_{\odot}$  & $n_{cz}$ & $1/gh_n$ & $h_n/R_{\odot}$  \\ \hline

0 & $1.18 \times 10^{-15}$ & 0.2359 & 0 & $-6.12 \times 10^{-12}$ & $-8.58 \times 10^{-5}$ \\ \hline
1 & $2.74 \times 10^{-15}$ & 0.1012 & 1 & $-2.09 \times 10^{-11}$ & $-2.51 \times 10^{-5}$ \\ \hline
2 & $5.85 \times 10^{-15}$ & 0.0474 & 2 & $-4.36 \times 10^{-11}$ & $-1.20 \times 10^{-5}$ \\ \hline
3 & $1.07 \times 10^{-14}$ & 0.0260 & 3 & $-7.36 \times 10^{-11}$ & $-7.14 \times 10^{-6}$ \\ \hline
4 & $1.71 \times 10^{-14}$ & 0.0162 & 4 & $-1.10 \times 10^{-10}$ & $-4.76 \times 10^{-6}$ \\ \hline
5 & $2.51 \times 10^{-14}$ & 0.0110 & 5 & $-1.54 \times 10^{-10}$ & $-3.42 \times 10^{-6}$ \\ \hline
6 & $3.49 \times 10^{-14}$ & 0.0080 & 6 & $-2.04 \times 10^{-10}$ & $-2.58 \times 10^{-6}$ \\ \hline
7 & $4.64 \times 10^{-14}$ & 0.0060 & 7 & $-2.60 \times 10^{-10}$ & $-2.02 \times 10^{-6}$ \\ \hline
8 & $5.95 \times 10^{-14}$ & 0.0047 & 8 & $-3.24 \times 10^{-10}$ & $-1.62 \times 10^{-6}$ \\ \hline
9 & $7.44 \times 10^{-14}$ & 0.0037 & 9 & $-3.94 \times 10^{-10}$ & $-1.33 \times 10^{-6}$ \\ \hline
10 & $9.10 \times 10^{-14}$ & 0.0030 & 10 & $-4.71 \times 10^{-10}$ & $-1.11 \times 10^{-6}$ \\ \hline

\end{tabular}

\caption{The two families of vertical eigenvalues when \cref{eq:vert} is solved for Model S's radial profile. The effective depth $h$ is calculated with $g = 5.18 \times 10^{4}$---the value at the base of the convection zone---for the radiative interior modes and $g = 2.74 \times 10^{4}$---the value at the surface---for the convection zone modes.}
\label{tab:eigenvalues}

\end{table*}
}

%% file: Bibliography.bib
@ARTICLE{Alshehhi2019,
       author = {{Alshehhi}, Rasha and {Hanson}, Chris S. and {Gizon}, Laurent and {Hanasoge}, Shravan},
        title = "{Supervised neural networks for helioseismic ring-diagram inversions}",
      journal = {\aap},
         year = 2019,
       volume = {622},
          eid = {A124},
        pages = {A124},
          doi = {10.1051/0004-6361/201834237},
archivePrefix = {arXiv},
       eprint = {1901.01505},
 primaryClass = {astro-ph.SR},
       adsurl = {https://ui.adsabs.harvard.edu/abs/2019A&A...622A.124A}
}

@ARTICLE{1988BCrAO..80..151G,
       author = {{Gorkin}, L.~B. and {Kosovichev}, A.~G.},
        title = "{Spatial structure of normal-mode solar oscillations}",
      journal = {Bulletin Crimean Astrophysical Observatory},
         year = 1988,
        month = jan,
       volume = {80},
        pages = {151},
       adsurl = {https://ui.adsabs.harvard.edu/abs/1988BCrAO..80..151G}
}

@book{Eckart1960,
    author = {{Eckart}, Carl},
    title = "{Hydrodynamics of Oceans and Atmospheres}",
    publisher = {Pergamon Press},
    year = 1960
}

@ARTICLE{1978A&A....70..597B,
       author = {{Berthomieu}, G. and {Gonczi}, G. and {Graff}, Ph. and {Provost}, J. and {Rocca}, A.},
        title = "{Low-frequency Gravity Modes of a Rotating Star}",
      journal = {\aap},
         year = 1978,
        month = nov,
       volume = {70},
        pages = {597},
       adsurl = {https://ui.adsabs.harvard.edu/abs/1978A&A....70..597B}
}

@ARTICLE{2008RvGeo..46.2004G,
       author = {{Gerkema}, T. and {Zimmerman}, J.~T.~F. and {Maas}, L.~R.~M. and {van Haren}, H.},
        title = "{Geophysical and astrophysical fluid dynamics beyond the traditional approximation}",
      journal = {Reviews of Geophysics},
         year = 2008,
        month = may,
       volume = {46},
       number = {2},
          eid = {RG2004},
        pages = {RG2004},
          doi = {10.1029/2006RG000220},
       adsurl = {https://ui.adsabs.harvard.edu/abs/2008RvGeo..46.2004G}
}

@ARTICLE{1995QJRMS.121..399W,
       author = {{White}, A.~A. and {Bromley}, R.~A.},
        title = "{Dynamically consistent, quasi-hydrostatic equations for global models with a complete representation of the Coriolis force}",
      journal = {Quarterly Journal of the Royal Meteorological Society},
         year = 1995,
        month = jan,
       volume = {121},
       number = {522},
        pages = {399-418},
          doi = {10.1002/qj.49712152208},
       adsurl = {https://ui.adsabs.harvard.edu/abs/1995QJRMS.121..399W}
}

@ARTICLE{2005QJRMS.131.2081W,
       author = {{White}, A.~A. and {Hoskins}, B.~J. and {Roulstone}, I. and {Staniforth}, A.},
        title = "{Consistent approximate models of the global atmosphere: shallow, deep, hydrostatic, quasi-hydrostatic and non-hydrostatic}",
      journal = {Quarterly Journal of the Royal Meteorological Society},
         year = 2005,
        month = jul,
       volume = {131},
       number = {609},
        pages = {2081-2107},
          doi = {10.1256/qj.04.49},
       adsurl = {https://ui.adsabs.harvard.edu/abs/2005QJRMS.131.2081W}
}

@ARTICLE{2025JFM..1007A..61P,
       author = {{Pedlosky}, Joseph},
        title = "{The dynamics of a thick fluid layer with a tilted rotation vector}",
      journal = {Journal of Fluid Mechanics},
         year = 2025,
        month = mar,
       volume = {1007},
          eid = {A61},
        pages = {A61},
          doi = {10.1017/jfm.2025.60},
       adsurl = {https://ui.adsabs.harvard.edu/abs/2025JFM..1007A..61P}
}

@ARTICLE{1997JGR...102.5733M,
       author = {{Marshall}, John and {Hill}, Chris and {Perelman}, Lev and {Adcroft}, Alistair},
        title = "{Hydrostatic, quasi-hydrostatic, and nonhydrostatic ocean modeling}",
      journal = {\jgr},
         year = 1997,
        month = mar,
       volume = {102},
       number = {C3},
        pages = {5733-5752},
          doi = {10.1029/96JC02776},
       adsurl = {https://ui.adsabs.harvard.edu/abs/1997JGR...102.5733M}
}

@ARTICLE{2021AA...652L...6G,
       author = {{Gizon}, Laurent and {Cameron}, Robert H. and {Bekki}, Yuto and {Birch}, Aaron C. and {Bogart}, Richard S. and {Brun}, Allan Sacha and {Damiani}, Cilia and {Fournier}, Damien and {Hyest}, Laura and {Jain}, Kiran and {Lekshmi}, B. and {Liang}, Zhi-Chao and {Proxauf}, Bastian},
        title = "{Solar inertial modes: Observations, identification, and diagnostic promise}",
      journal = {\aap},
         year = 2021,
        month = aug,
       volume = {652},
          eid = {L6},
        pages = {L6},
          doi = {10.1051/0004-6361/202141462},
archivePrefix = {arXiv},
       eprint = {2107.09499},
 primaryClass = {astro-ph.SR},
       adsurl = {https://ui.adsabs.harvard.edu/abs/2021A&A...652L...6G}
}

@ARTICLE{Hanasoge2019,
       author = {{Hanasoge}, Shravan and {Mandal}, Krishnendu},
        title = "{Detection of Rossby Waves in the Sun using Normal-mode Coupling}",
      journal = {\apjl},
         year = 2019,
        month = feb,
       volume = {871},
       number = {2},
          eid = {L32},
        pages = {L32},
          doi = {10.3847/2041-8213/aaff60},
       adsurl = {https://ui.adsabs.harvard.edu/abs/2019ApJ...871L..32H}
}

@ARTICLE{Hanson2020,
       author = {{Hanson}, Chris S. and {Gizon}, Laurent and {Liang}, Zhi-Chao},
        title = "{Solar Rossby waves observed in GONG++ ring-diagram flow maps}",
      journal = {\aap},
         year = 2020,
        month = mar,
       volume = {635},
          eid = {A109},
        pages = {A109},
          doi = {10.1051/0004-6361/201937321},
archivePrefix = {arXiv},
       eprint = {2002.01194},
 primaryClass = {astro-ph.SR},
       adsurl = {https://ui.adsabs.harvard.edu/abs/2020A&A...635A.109H}
}

@ARTICLE{2024ApJ...967...46M,
       author = {{Mandal}, Krishnendu and {Hanasoge}, Shravan M.},
        title = "{Probing Depth Variations of Solar Inertial Modes through Normal Mode Coupling}",
      journal = {\apj},
         year = 2024,
        month = may,
       volume = {967},
       number = {1},
          eid = {46},
        pages = {46},
          doi = {10.3847/1538-4357/ad391b},
archivePrefix = {arXiv},
       eprint = {2403.08150},
 primaryClass = {astro-ph.SR},
       adsurl = {https://ui.adsabs.harvard.edu/abs/2024ApJ...967...46M}
}

@ARTICLE{Hathaway2021,
       author = {{Hathaway}, David H. and {Upton}, Lisa A.},
        title = "{Hydrodynamic Properties of the Sun's Giant Cellular Flows}",
      journal = {\apj},
         year = 2021,
        month = feb,
       volume = {908},
       number = {2},
          eid = {160},
        pages = {160},
          doi = {10.3847/1538-4357/abcbfa},
archivePrefix = {arXiv},
       eprint = {2006.06084},
 primaryClass = {astro-ph.SR},
       adsurl = {https://ui.adsabs.harvard.edu/abs/2021ApJ...908..160H}
}

@ARTICLE{1940TrAGU..21..262H,
       author = {{Haurwitz}, B.},
        title = "{Atmospheric disturbances on the rotating Earth}",
      journal = {Transactions, American Geophysical Union},
         year = 1940,
        month = jan,
       volume = {21},
       number = {2},
        pages = {262-264},
          doi = {10.1029/TR021i002p00262},
       adsurl = {https://ui.adsabs.harvard.edu/abs/1940TrAGU..21..262H}
}

@ARTICLE{2020A&A...637A..65D,
       author = {{Damiani}, C. and {Cameron}, R.~H. and {Birch}, A.~C. and {Gizon}, L.},
        title = "{Rossby modes in slowly rotating stars: depth dependence in distorted polytropes with uniform rotation}",
      journal = {\aap},
         year = 2020,
        month = may,
       volume = {637},
          eid = {A65},
        pages = {A65},
          doi = {10.1051/0004-6361/201936251},
archivePrefix = {arXiv},
       eprint = {2003.05276},
 primaryClass = {astro-ph.SR},
       adsurl = {https://ui.adsabs.harvard.edu/abs/2020A&A...637A..65D}
}

@ARTICLE{1996Sci...272.1286C,
       author = {{Christensen-Dalsgaard}, J. and {Dappen}, W. and {Ajukov}, S.~V. and {Anderson}, E.~R. and {Antia}, H.~M. and {Basu}, S. and {Baturin}, V.~A. and {Berthomieu}, G. and {Chaboyer}, B. and {Chitre}, S.~M. and {Cox}, A.~N. and {Demarque}, P. and {Donatowicz}, J. and {Dziembowski}, W.~A. and {Gabriel}, M. and {Gough}, D.~O. and {Guenther}, D.~B. and {Guzik}, J.~A. and {Harvey}, J.~W. and {Hill}, F. and {Houdek}, G. and {Iglesias}, C.~A. and {Kosovichev}, A.~G. and {Leibacher}, J.~W. and {Morel}, P. and {Proffitt}, C.~R. and {Provost}, J. and {Reiter}, J. and {Rhodes}, Jr., E.~J. and {Rogers}, F.~J. and {Roxburgh}, I.~W. and {Thompson}, M.~J. and {Ulrich}, R.~K.},
        title = "{The Current State of Solar Modeling}",
      journal = {Science},
         year = 1996,
        month = may,
       volume = {272},
       number = {5266},
        pages = {1286-1292},
          doi = {10.1126/science.272.5266.1286},
       adsurl = {https://ui.adsabs.harvard.edu/abs/1996Sci...272.1286C}
}

@ARTICLE{2019A&A...626A...3L,
       author = {{Liang}, Zhi-Chao and {Gizon}, Laurent and {Birch}, Aaron C. and {Duvall}, Thomas L.},
        title = "{Time-distance helioseismology of solar Rossby waves}",
      journal = {\aap},
         year = 2019,
        month = jun,
       volume = {626},
          eid = {A3},
        pages = {A3},
          doi = {10.1051/0004-6361/201834849},
archivePrefix = {arXiv},
       eprint = {1812.07413},
 primaryClass = {astro-ph.SR},
       adsurl = {https://ui.adsabs.harvard.edu/abs/2019A&A...626A...3L}
}

@ARTICLE{Loptien2018,
       author = {{L{\"o}ptien}, Bj{\"o}rn and {Gizon}, Laurent and {Birch}, Aaron C. and {Schou}, Jesper and {Proxauf}, Bastian and {Duvall}, Thomas L. and {Bogart}, Richard S. and {Christensen}, Ulrich R.},
        title = "{Global-scale equatorial Rossby waves as an essential component of solar internal dynamics}",
      journal = {Nature Astronomy},
         year = 2018,
        month = may,
       volume = {2},
        pages = {568-573},
          doi = {10.1038/s41550-018-0460-x},
archivePrefix = {arXiv},
       eprint = {1805.07244},
 primaryClass = {astro-ph.SR},
       adsurl = {https://ui.adsabs.harvard.edu/abs/2018NatAs...2..568L}
}

@ARTICLE{Proxauf2020,
       author = {{Proxauf}, B. and {Gizon}, L. and {L{\"o}ptien}, B. and {Schou}, J. and {Birch}, A.~C. and {Bogart}, R.~S.},
        title = "{Exploring the latitude and depth dependence of solar Rossby waves using ring-diagram analysis}",
      journal = {\aap},
         year = 2020,
        month = feb,
       volume = {634},
          eid = {A44},
        pages = {A44},
          doi = {10.1051/0004-6361/201937007},
archivePrefix = {arXiv},
       eprint = {1912.02056},
 primaryClass = {astro-ph.SR},
       adsurl = {https://ui.adsabs.harvard.edu/abs/2020A&A...634A..44P}
}

@article{Rossby1939RelationBV,
  title={Relation between variations in the intensity of the zonal circulation of the atmosphere and the displacements of the semi-permanent centers of action},
  author={C-G. Rossby},
  journal={Journal of Marine Research},
  year={1939},
  volume={2},
  pages={38-55}
}

@ARTICLE{1986SoPh..105....1W,
       author = {{Wolff}, C.~L. and {Blizard}, J.~B.},
        title = "{Properties of R-Modes in the Sun}",
      journal = {\solphys},
         year = 1986,
        month = may,
       volume = {105},
       number = {1},
        pages = {1-15},
          doi = {10.1007/BF00156371},
       adsurl = {https://ui.adsabs.harvard.edu/abs/1986SoPh..105....1W}
}

@ARTICLE{2023A&A...671A..91A,
       author = {{Albekioni}, M. and {Zaqarashvili}, T.~V. and {Kukhianidze}, V.},
        title = "{Rossby waves on stellar equatorial {\ensuremath{\beta}} planes: Uniformly rotating radiative stars}",
      journal = {\aap},
         year = 2023,
        month = mar,
       volume = {671},
          eid = {A91},
        pages = {A91},
          doi = {10.1051/0004-6361/202243985},
archivePrefix = {arXiv},
       eprint = {2301.07446},
 primaryClass = {astro-ph.SR},
       adsurl = {https://ui.adsabs.harvard.edu/abs/2023A&A...671A..91A}
}

@ARTICLE{2022AA...662A..16B,
       author = {{Bekki}, Yuto and {Cameron}, Robert H. and {Gizon}, Laurent},
        title = "{Theory of solar oscillations in the inertial frequency range: Linear modes of the convection zone}",
      journal = {\aap},
         year = 2022,
        month = jun,
       volume = {662},
          eid = {A16},
        pages = {A16},
          doi = {10.1051/0004-6361/202243164},
archivePrefix = {arXiv},
       eprint = {2203.04442},
 primaryClass = {astro-ph.SR},
       adsurl = {https://ui.adsabs.harvard.edu/abs/2022A&A...662A..16B}
}

@ARTICLE{1978MNRAS.182..423P,
       author = {{Papaloizou}, J. and {Pringle}, J.~E.},
        title = "{Non-radial oscillations of rotating stars and their relevance to the short-period oscillations of cataclysmic variables.}",
      journal = {\mnras},
         year = 1978,
        month = feb,
       volume = {182},
        pages = {423-442},
          doi = {10.1093/mnras/182.3.423},
       adsurl = {https://ui.adsabs.harvard.edu/abs/1978MNRAS.182..423P}
}

@ARTICLE{1981A&A....94..126P,
       author = {{Provost}, J. and {Berthomieu}, G. and {Rocca}, A.},
        title = "{Low Frequency Oscillations of a Slowly Rotating Star - Quasi Toroidal Modes}",
      journal = {\aap},
         year = 1981,
        month = jan,
       volume = {94},
        pages = {126},
       adsurl = {https://ui.adsabs.harvard.edu/abs/1981A&A....94..126P}
}

@ARTICLE{2022AA...666A.135B,
       author = {{Bekki}, Yuto and {Cameron}, Robert H. and {Gizon}, Laurent},
        title = "{Theory of solar oscillations in the inertial frequency range: Amplitudes of equatorial modes from a nonlinear rotating convection simulation}",
      journal = {\aap},
         year = 2022,
        month = oct,
       volume = {666},
          eid = {A135},
        pages = {A135},
          doi = {10.1051/0004-6361/202244150},
archivePrefix = {arXiv},
       eprint = {2208.11081},
 primaryClass = {astro-ph.SR},
       adsurl = {https://ui.adsabs.harvard.edu/abs/2022A&A...666A.135B}
}

@ARTICLE{Waidele2023,
       author = {{Waidele}, M. and {Zhao}, Junwei},
        title = "{Observed Power and Frequency Variations of Solar Rossby Waves with Solar Cycles}",
      journal = {\apjl},
         year = 2023,
        month = sep,
       volume = {954},
       number = {1},
          eid = {L26},
        pages = {L26},
          doi = {10.3847/2041-8213/acefd0},
archivePrefix = {arXiv},
       eprint = {2308.07040},
 primaryClass = {astro-ph.SR},
       adsurl = {https://ui.adsabs.harvard.edu/abs/2023ApJ...954L..26W}
}

@book{Ince1956,
  title     = "Ordinary Differential Equations",
  author    = "Ince, E.L.",
  year      = 1956,
  publisher = "Dover Publications, Inc.",
  address   = "USA"
}

@ARTICLE{1968RSPTA.262..511L,
       author = {{Longuet-Higgins}, M.~S.},
        title = "{The Eigenfunctions of Laplace's Tidal Equations over a Sphere}",
      journal = {Philosophical Transactions of the Royal Society of London Series A},
         year = 1968,
        month = feb,
       volume = {262},
       number = {1132},
        pages = {511-607},
          doi = {10.1098/rsta.1968.0003},
       adsurl = {https://ui.adsabs.harvard.edu/abs/1968RSPTA.262..511L}
}

@BOOK{2017aofd.book.....V,
       author = {{Vallis}, Geoffrey K.},
        title = "{Atmospheric and Oceanic Fluid Dynamics: Fundamentals and Large-Scale Circulation}",
         year = 2017,
        publisher = {Cambridge University Press},
          doi = {10.1017/9781107588417},
       adsurl = {https://ui.adsabs.harvard.edu/abs/2017aofd.book.....V}
}

@BOOK{Pedlosky1987,
       author = {{Pedlosky}, Joseph},
        title = "{Geophysical Fluid Dynamics}",
        publisher={Springer-Verlag New York Inc},
         year = 1987
}

@BOOK{1983tsp..book.....C,
       author = {{Cox}, J.~P.},
        title = "{Theory of stellar pulsations.}",
         year = 1983,
        publisher = {Princeton University Press},
       adsurl = {https://ui.adsabs.harvard.edu/abs/1983tsp..book.....C}
}

@ARTICLE{Hanson2024,
       author = {{Hanson}, C.~S. and {Hanasoge}, S.},
        title = "{Existence of small-scale Rossby waves points to low convective velocity amplitudes in the Sun}",
      journal = {Physics of Fluids},
         year = 2024,
        month = aug,
       volume = {36},
       number = {8},
          eid = {086626},
        pages = {086626},
          doi = {10.1063/5.0216403},
       adsurl = {https://ui.adsabs.harvard.edu/abs/2024PhFl...36h6626H}
}

@ARTICLE{1949MSRSL...9....3L,
       author = {{Ledoux}, P.},
        title = "{Contributions {\`a} l'Etude de la Structure Interne des Etoiles et de leur Stabilit{\'e}}",
      journal = {Memoires of the Societe Royale des Sciences de Liege},
         year = 1949,
        month = jan,
       volume = {9},
        pages = {3-294},
       adsurl = {https://ui.adsabs.harvard.edu/abs/1949MSRSL...9....3L}
}

@ARTICLE{JCD1996,
       author = {{Christensen-Dalsgaard}, J. and {Dappen}, W. and {Ajukov}, S.~V. and {Anderson}, E.~R. and {Antia}, H.~M. and {Basu}, S. and {Baturin}, V.~A. and {Berthomieu}, G. and {Chaboyer}, B. and {Chitre}, S.~M. and {Cox}, A.~N. and {Demarque}, P. and {Donatowicz}, J. and {Dziembowski}, W.~A. and {Gabriel}, M. and {Gough}, D.~O. and {Guenther}, D.~B. and {Guzik}, J.~A. and {Harvey}, J.~W. and {Hill}, F. and {Houdek}, G. and {Iglesias}, C.~A. and {Kosovichev}, A.~G. and {Leibacher}, J.~W. and {Morel}, P. and {Proffitt}, C.~R. and {Provost}, J. and {Reiter}, J. and {Rhodes}, Jr., E.~J. and {Rogers}, F.~J. and {Roxburgh}, I.~W. and {Thompson}, M.~J. and {Ulrich}, R.~K.},
        title = "{The Current State of Solar Modeling}",
      journal = {Science},
         year = 1996,
        month = may,
       volume = {272},
       number = {5266},
        pages = {1286-1292},
          doi = {10.1126/science.272.5266.1286},
       adsurl = {https://ui.adsabs.harvard.edu/abs/1996Sci...272.1286C}
}

@ARTICLE{jcd2002,
       author = {{Christensen-Dalsgaard}, J{\o}rgen},
        title = "{Helioseismology}",
      journal = {Reviews of Modern Physics},
         year = 2002,
        month = nov,
       volume = {74},
       number = {4},
        pages = {1073-1129},
          doi = {10.1103/RevModPhys.74.1073},
archivePrefix = {arXiv},
       eprint = {astro-ph/0207403},
 primaryClass = {astro-ph},
       adsurl = {https://ui.adsabs.harvard.edu/abs/2002RvMP...74.1073C}
}

@INPROCEEDINGS{Gough1993,
       author = {{Gough}, D.~O.},
        title = "{Linear adiabatic stellar pulsation.}",
    booktitle = {Astrophysical Fluid Dynamics - Les Houches 1987},
         year = 1993,
       editor = {{Zahn}, J. -P. and {Zinn-Justin}, J.},
        month = jan,
        pages = {399-560},
       adsurl = {https://ui.adsabs.harvard.edu/abs/1993afd..conf..399G}
}

@ARTICLE{Hindman2022,
       author = {{Hindman}, Bradley W. and {Jain}, Rekha},
        title = "{Radial Trapping of Thermal Rossby Waves within the Convection Zones of Low-mass Stars}",
      journal = {\apj},
         year = 2022,
        month = jun,
       volume = {932},
       number = {1},
          eid = {68},
        pages = {68},
          doi = {10.3847/1538-4357/ac6d64},
archivePrefix = {arXiv},
       eprint = {2205.02346},
 primaryClass = {astro-ph.SR},
       adsurl = {https://ui.adsabs.harvard.edu/abs/2022ApJ...932...68H}
}

@ARTICLE{Hindman2024,
       author = {{Hindman}, Bradley W. and {Julien}, Keith},
        title = "{Low-frequency Internal Gravity Waves Are Pseudo-incompressible}",
      journal = {\apj},
         year = 2024,
        month = jan,
       volume = {960},
       number = {1},
          eid = {64},
        pages = {64},
          doi = {10.3847/1538-4357/ad0967},
archivePrefix = {arXiv},
       eprint = {2309.10079},
 primaryClass = {physics.flu-dyn},
       adsurl = {https://ui.adsabs.harvard.edu/abs/2024ApJ...960...64H}
}
